\documentclass{article}
\usepackage[utf8]{inputenc}
\usepackage{graphicx}
\usepackage{times}
\usepackage{xcolor}
\usepackage{amsmath}
\usepackage{subfigure}
\usepackage{amssymb}
\usepackage{epstopdf}
\usepackage[numbers,sort&compress]{natbib}
\usepackage{appendix}
\usepackage{multirow}
\newcommand{\be}{\begin{equation}
\newcommand{\ee}{\end{equation}}}
\newcommand{\bea}{\begin{eqnarray}}
\newcommand{\eea}{\end{eqnarray}}
\newcommand{\nn}{\nonumber}

\begin{document}

\title{The Treatment of conformable systems with second class constraints  }

\author{Eqab.M.Rabei$^1$, Mohamed.Al-Masaeed $^2$, and Dumitru Baleanu$^3$\\
$^1$Physics Department, Faculty of Science, Al al-Bayt University,\\ P.O. Box 130040, Mafraq 25113, Jordan\\ $^2$Ministry of Education, Jordan\\$^3$Department of Mathematics, Cankaya University, Ankara, Turkey\\eqabrabei@gmail.com\\moh.almssaeed@gmail.com\\dumitru@cankaya.edu.tr}

\maketitle


\begin{abstract}
In this work, a conformable singular system with second-class constraints is discussed. The conformable Poisson  bracket    (CDB) of two functions is defined. and, the Dirac theory is developed to be applicable to conformable singular systems. In addition, to demonstrate the theory an illustrative example is given and solved, the results are found to be in agreement with ref \cite{rabei1992hamilton} when $\alpha=1$.
\\

\end{abstract}

\section{Introduction}

constrained systems, also known as Singular systems, are a prominent framework of Physics. When dealing with physical systems subject to certain constraints, these constraints impose limitations on the degrees of freedom, affecting the dynamics and behavior of the system, etc \cite{dirac_lectures_1966,rabei1992hamilton,PhysRevA.66.024101,guler1989dynamics}. In other words, the constraints impose additional relations on the system's variables, resulting in a reduction of the number of independent degrees of freedom. This reduction leads to unique phenomena in understanding the singular systems.\\
Singular systems find applications in various branches of physics, including classical mechanic quantum mechanics and field theory \cite{rabei1999canonical,brown_singular_2022,rabei_quantization_2003,brown_singular_2023,eshraim2023treatment}.\\
In the Last few decades, fractional Calculus and physical applications have been developed. Particularly in Lagrangian and Hamiltonian formulations of non-conservative systems, fractional calculus has been applied. \cite{rabei2004potentials,rabei2004hamiltonian,rabei2007hamilton,lazo2016variational}.
The field of mathematics known as fractional calculus focuses on applying non-integer orders to differentiation and integration.
This powerful mathematical framework has found wide applications in science and engineering. Due to its capacity to offer a more exact description of different physical phenomena, this field has attracted a lot of interest recently: There are numerous ways to define fractional integral and derivative. \cite{oldhamFractionalCalculusTheory1974,millerIntroductionFractionalIntegrals1993,kilbas2006theory}.\\
In \cite{khalil2014new} Khalil et al proposed a new definition which is
called a conformable derivative, it has common properties with the known traditional derivative. For a given function
 $f(t)\in [0,\infty) \to {R}$, the conformable derivative of $f(t)$ of order $\alpha $ where $0 < \alpha \leq 1$, denoted as $D_{t}^\alpha f(t)$ is defined as \cite{khalil2014new}:
\be
\label{conformable}
D_{t}^\alpha f(t)=\lim_{\epsilon \to 0}\frac{f(t+\epsilon t^{1-\alpha})-f(t)}{\epsilon}= t^{1-\alpha} \frac{d}{dt}f(t).
\ee
In \cite{abdeljawad2015conformable,atangana2015new,jarad2017new}, this CFD is re-investigated and new properties similar to these in traditional calculus were derived and discussed.
Conformable calculus has numerous applications in many fields of physics such as quantum mechanics \cite{mozaffari2018investigation,al2021quantization,chung2020effect,al2022extension,al2021extension,al2022wkb,albanwa_quantization_2023}, nuclear physics \cite{HAMMAD2021122307,HAMMAD2021122203}, special relativity \cite{al-jamel_effect_2022,pawar2018approach}, classical mechanics \cite{chung2021new,rabei2018quantization} and mathematical physics \cite{rabei_solution_2021,shihab2021associated,hammad2014conformable}. 
\section{Conformable singular system }
Let us define the conformable Lagrangian in the following form 
\be
L(t,D^{\alpha-1}q_i,D^\alpha q_i) = a_{ij} D^\alpha q_iD^\alpha q_j + b_j D^\alpha q_j + V(D^{\alpha-1} q_i),
\ee
where $0< \alpha\leq 1$.\\
Noting that $D^\alpha q_i$ is the conformable derivative of the coordinate $ q_i$ which represented the conformable velocity and $D^{\alpha-1}q_i$ is the canonical conformable conjugate coordinate. 
The Hessian matrix for the conformable Lagrangian is defined as 
\be
\label{hm}
W_{ij}= \frac{\partial^2 L}{\partial D^\alpha q_i \partial D^\alpha q_j}, \quad i,j=1,2,\dots, n.
\ee
If this matrix has   rank  $n$ the system is called regular and can be treated using traditional conformable 
 mechanics and the conformable regular systems have been treated by many authors \cite{lazo2016variational} and references therein, while systems with a rank less than $n$ are called conformable  singular systems. Traditional singular systems were treated using two methods: the Dirac method \cite{dirac_lectures_1966} and the canonical method \cite{rabei1992hamilton}.\\
The conformable singular systems are discussed by Rabei and Horani \cite{rabei2018quantization} using the canonical method. The conformable  systems which have the rank of the Hessian matrix less than $n$ are called conformable   constrained systems. Let us assume that the rank of the Hessian matrix is $r$ which is less than $n$. Following  to Dirac \cite{dirac_lectures_1966}, this implies the existence of $(n-r)$ conformable primary constraints. Thus, we may define the total conformable 
 Hamiltonian as
\be
\label{totalh}
H_{T\alpha}= H_{0\alpha} + \nu_\mu H^{'}_{\mu\alpha}, \quad \mu=n-r+1, \dots,n.
\ee
and $ H_{0\alpha}$ being the conformable Hamiltonian
\be
\label{h0}
H_{0\alpha}  = -L (t,D^{\alpha-1} q_i,D^\alpha q_i) + P_{i\alpha} D^\alpha q_i, \quad i=1,2,\dots,n.
\ee
Where $P_{i\alpha}$ is the conformable generalized momenta  defined as 
\be
P_{a\alpha} = \frac{\partial}{\partial(D^\alpha q_a)}, \quad a=1,2,\dots,n-r.
\ee
\be
\label{pmu}
P_{\mu \alpha} = \frac{\partial}{\partial(D^\alpha q_a)}, \quad \mu=n-p+1,2,\dots,n.
\ee
$ \nu_\mu$  are unknown coefficients and $H^{'}_{\mu\alpha}$ are the conformable primary constraints which can be obtained using eq.\eqref{pmu} as follow 
\be
\label{eq8}
H^{'}_{\mu\alpha} = P_{\mu \alpha} + H_\mu (D^{\alpha-1} q_i,D^\alpha q_i,P_{a\alpha}), \quad \nu=n-p+1,\dots,n.
\ee
The conformable Poisson bracket (CPB)  of two functions $f(D^{\alpha-1} q_i,P_{a\alpha})$ and  
\\$g(D^{\alpha-1} q_i,P_{a\alpha})$ is defined as 
\be
\label{CPB}
\{f,g\}_\alpha = \frac{\partial f}{\partial(D^{\alpha-1} q_i)}\frac{\partial g}{\partial P_{i\alpha}}-\frac{\partial f }{\partial P_{i\alpha}}\frac{\partial g}{\partial(D^{\alpha-1} q_i)}
\ee
The total time derivative of any function $g$ in terms of the generalized coordinates $D^{\alpha-1} q_i$ and momenta $ P_{i\alpha}$ is given as 
\be
\label{gdot}
\Dot{g} = \{g,H_{0\alpha} \}_\alpha+ \nu_\mu \{g,H_{\mu\alpha}\}_\alpha.
\ee
according to Dirac \cite{dirac_lectures_1966}, according to the  consistency conditions the total time derivative of the primary constraints $H^\prime_\mu$ should be equal to zero. Thus, 
\be
\label{hdot}
\Dot{H^\prime_{\mu\alpha}} = \{H^\prime_{\mu\alpha} , H_{0\alpha}\}_\alpha + \nu_\mu \{H^\prime_{\mu\alpha} , H^\prime_{\nu\alpha} \}_\alpha \approx 0.
\ee
This equation may be solved to obtain the unknowns $\nu_\mu$ or lead to new constraints that restrict the motion \cite{dirac_lectures_1966}.
Following Dirac, the constraints which have vanishing CPB's are called Conformable  First Class and which do not have vanishing CPB's are  called Conformable Second Class constraints and the equations of motion will be proposed as,
\be
D^\alpha q_i=\{D^{\alpha-1} q_i,H_{T\alpha}\}_\alpha
\ee
\be
D P_{i\alpha}=\{ P_{i\alpha},H_{T\alpha}\}_\alpha
\ee

\section{Conformable Second Class Constraints}
To demonstrate our theory we would like to give a model of the conformable constraints of   the Second Class  \cite{rabei1992hamilton}. Investigate the following Lagrangian 
\bea
\label{lag}
L&=& \frac{1}{2}(D^\alpha q_1)^2 - \frac{1}{4}((D^\alpha q_2)^2-2 D^\alpha q_2 D^\alpha q_3+(D^\alpha q_3)^2)
\\\nn&+& (D^{\alpha-1} q_1 +D^{\alpha-1} q_3)D^\alpha q_2 - (D^{\alpha-1} q_1 +D^{\alpha-1} q_2 + (D^{\alpha-1} q_3)^2).
\eea
The Hessian matrix $W_{33}$ can be constructed. 
\be
W_{33}= \begin{bmatrix}
   \frac{\partial^2 L }{\partial D^\alpha q_1\partial D^\alpha q_1}  & \frac{\partial^2 L }{\partial D^\alpha q_1\partial D^\alpha q_2}&\frac{\partial^2 L }{\partial D^\alpha q_1\partial D^\alpha q_3} \\[0.3em]
     \frac{\partial^2 L }{\partial D^\alpha q_2\partial D^\alpha q_1}&\frac{\partial^2 L }{\partial D^\alpha q_2\partial D^\alpha q_3}&\frac{\partial^2 L }{\partial D^\alpha q_2\partial D^\alpha q_3} \\[0.3em]
     \frac{\partial^2 L }{\partial D^\alpha q_3\partial D^\alpha q_1}& \frac{\partial^2 L }{\partial D^\alpha q_3\partial D^\alpha q_2}& \frac{\partial^2 L }{\partial D^\alpha q_3\partial D^\alpha q_3}
\end{bmatrix}
\ee
It is easy to show that the rank of this matrix is two. Then, the momenta read as,
\bea
\label{p1}
P_{1\alpha}&=& \frac{\partial L}{\partial(D^\alpha q_1)}=D^\alpha q_1,
\\\label{p2} P_{2\alpha}&=& \frac{\partial L}{\partial(D^\alpha q_2)}= - \frac{1}{2} D^\alpha q_2 + \frac{1}{2} D^\alpha q_3 + (D^{\alpha-1} q_1 +D^{\alpha-1} q_3),
\\\label{p3} P_{3\alpha}&=& \frac{\partial L}{\partial(D^\alpha q_3)}= \frac{1}{2} D^\alpha q_2 - \frac{1}{2} D^\alpha q_3.
\eea
Thus, the Hamiltonian $H_0$ can be calculated as 
\be
H_{0\alpha}= \frac{1}{2} P^2_\alpha - P^2_{3\alpha}+ D^{\alpha-1} q_1 +D^{\alpha-1} q_2 +(D^{\alpha-1} q_3)^2.
\ee
Substituting eq.\eqref{p3} in eq.\eqref{p2} one may obtain 
\be
P_{2\alpha}= - P_{3\alpha} + D^{\alpha-1} q_1 + D^{\alpha-1} q_2.
\ee 
The conformable momentum $P_{2\alpha}$ is not independent. Thus, following  to eq.\eqref{eq8}, these are conformable primary constraints which can be written as 
\be
\label{prim cons}
H^\prime_{2\alpha}= P_{2\alpha} + P_{3\alpha} -D^{\alpha-1} q_1 - D^{\alpha-1} q_2.
\ee 
Making use of equation \eqref{totalh}, the total conformable Hamiltonian reads as 
\bea
\nn
H_{T\alpha} &=& \frac{1}{2}  P_{\alpha}^2- P_{3\alpha}^2+D^{\alpha-1} q_1 + D^{\alpha-1} q_2 + (D^{\alpha-1} q_3)^2
\\&+& \nu_2 (P_{2\alpha}+P_{3\alpha}-D^{\alpha-1} q_1 - D^{\alpha-1} q_2).
\eea
Now, the total time derivative of the conformable primary constraints should be equal to zero. So, using eq.\eqref{gdot} we obtain 
\bea
\Dot{H}^\prime_{2\alpha} &=& \{H^\prime_{2\alpha},H_T\}_\alpha =  \{H^\prime_{2\alpha},H_{0\alpha}\}_\alpha + \nu_2 \{H^\prime_{2\alpha},H^\prime_{2\alpha}\}_\alpha =0
\\\nn&=& \frac{\partial H^\prime_{2\alpha}}{\partial D^{\alpha-1} q_1}\frac{\partial H_{0\alpha}}{\partial P_{1\alpha}}
-\frac{\partial H_{0\alpha}}{\partial D^{\alpha-1} q_1} \frac{\partial H^\prime_{2\alpha}}{\partial P_{1\alpha}}
+\frac{\partial H^\prime_{2\alpha}}{\partial D^{\alpha-1} q_2} \frac{\partial H_{0\alpha}}{\partial P_{2\alpha}}
-\frac{\partial H^\prime_{2\alpha}}{\partial P_{2\alpha}} \frac{\partial H_{0\alpha}}{\partial D^{\alpha-1} q_2}
\\\nn&+&\frac{\partial H^\prime_{2\alpha}}{\partial D^{\alpha-1} q_3} \frac{\partial H_{0\alpha}}{\partial P_{3\alpha}}
-\frac{\partial H^\prime_{2\alpha}}{\partial P_{3\alpha}} \frac{\partial H_{0\alpha}}{\partial D^{\alpha-1} q_3}
\\\nn&=& P_{1\alpha}-1+2 P_{3\alpha}-2 D^{\alpha-1} q_3=0.
\eea
This consistency condition imposes a new constraint
\be
H^\prime_{1\alpha} = 2  P_{3\alpha} - P_{1\alpha} -2 D^{\alpha-1} q_3-1.
\ee
Again the total time derivative of the new constraint should be equal to  zero.
\bea
\Dot{H}^\prime_{1\alpha} &=& \{H^\prime_{1\alpha},H_T\}_\alpha =  \{H^\prime_{1\alpha},H_{0\alpha}\}_\alpha + \nu_2 \{H^\prime_{1\alpha},H^\prime_{2\alpha}\}_\alpha =0
\\\nn&=& \frac{\partial H^\prime_{1\alpha}}{\partial D^{\alpha-1} q_1}\frac{\partial H_{0\alpha}}{\partial P_{1\alpha}}
-\frac{\partial H_{0\alpha}}{\partial D^{\alpha-1} q_1} \frac{\partial H^\prime_{1\alpha}}{\partial P_{1\alpha}}
+\frac{\partial H^\prime_{1\alpha}}{\partial D^{\alpha-1} q_2} \frac{\partial H_{0\alpha}}{\partial P_{2\alpha}}
-\frac{\partial H^\prime_{1\alpha}}{\partial P_{2\alpha}} \frac{\partial H_{0\alpha}}{\partial D^{\alpha-1} q_2} 
\\\nn&+&\frac{\partial H^\prime_{1\alpha}}{\partial D^{\alpha-1} q_3} \frac{\partial H_{0\alpha}}{\partial P_{3\alpha}}
-\frac{\partial H^\prime_{1\alpha}}{\partial P_{3\alpha}} \frac{\partial H_{0\alpha}}{\partial D^{\alpha-1} q_3} 
+  \nu_2 ( \frac{\partial H^\prime_{1\alpha}}{\partial D^{\alpha-1} q_1}\frac{\partial H_{2\alpha}}{\partial P_{1\alpha}}
-\frac{\partial H_{2\alpha}}{\partial D^{\alpha-1} q_1} \frac{\partial H^\prime_{1\alpha}}{\partial P_{1\alpha}}
\\\nn&+&\frac{\partial H^\prime_{1\alpha}}{\partial D^{\alpha-1} q_2} \frac{\partial H_{2\alpha}}{\partial P_{2\alpha}}
-\frac{\partial H^\prime_{1\alpha}}{\partial P_{2\alpha}} \frac{\partial H_{2\alpha}}{\partial D^{\alpha-1} q_2} 
+\frac{\partial H^\prime_{1\alpha}}{\partial D^{\alpha-1} q_3} \frac{\partial H_{2\alpha}}{\partial P_{3\alpha}}
-\frac{\partial H^\prime_{1\alpha}}{\partial P_{3\alpha}} \frac{\partial H_{2\alpha}}{\partial D^{\alpha-1} q_3} )
\\\nn&=&1+4 P_{3\alpha}-4 D^{\alpha-1} q_3 - \nu_2 = 0.
\eea
Then, we arrive at the result 
\be
\nu_2 = - (4 D^{\alpha-1} q_3 -4 P_{3\alpha}-1).
\ee
There is no further constraint, and the unknown $\nu_2$ is determined and the conformable constraints $H^\prime_{2\alpha}$ and $H^\prime_{1\alpha}$ are second class. Thus , the total Hamiltonian read as 
\bea
H_{T\alpha} &=& \frac{1}{2}P_{\alpha}^2 - P_{3\alpha}^2+ D^{\alpha-1} q_1 +D^{\alpha-1} q_2 + (D^{\alpha-1} q_3 )^2\\\nn &-& (4 D^{\alpha-1} q_3 -4 P_{3\alpha}-1)(P_{2\alpha}+P_{3\alpha} -D^{\alpha-1} q_1 - D^{\alpha-1} q_3),
\eea
and the equation of motion can be calculated as 
\bea
\nn
D^{\alpha} q_1 &=& \{D^{\alpha-1} q_1,H_T\}_\alpha =  \{D^{\alpha-1} q_1,\frac{1}{2}P_{1\alpha}^2\}_\alpha = \frac{1}{2}P_{1\alpha}\{D^{\alpha-1} q_1,P_{1\alpha}\}_\alpha 
\\ \label{q1dot} &+&\{D^{\alpha-1} q_1,P_{1\alpha}\}_\alpha \frac{1}{2}P_{1\alpha}=P_{1\alpha}.
\eea
\bea
\nn
D^{\alpha} q_2 &=& \{D^{\alpha-1} q_2,H_T\}_\alpha =\nu_2  \{D^{\alpha-1} q_2,P_{2\alpha}\}_\alpha =\nu_2 
\\ \label{q2}&=&4P_{3\alpha} -4 D^{\alpha-1} q_3 +1.
\eea
\bea
\nn
D^{\alpha} q_3 &=& \{D^{\alpha-1} q_3,H_T\}_\alpha = -  \{D^{\alpha-1} q_3,P_{3\alpha}^2\}_\alpha
\\\nn&+&  \{D^{\alpha-1} q_3,4P_{3\alpha}\}_\alpha (P_{2\alpha}+P_{3\alpha}-D^{\alpha-1} q_1-D^{\alpha-1} q_3)
\\\nn&-&(4D^{\alpha-1} q_3 - 4 P_{3\alpha} -1 ) \{D^{\alpha-1} q_3,P_{3\alpha}\}_\alpha
\\\nn&=& -2P_{3\alpha} + 4(P_{2\alpha}+P_{3\alpha}-D^{\alpha-1} q_1-D^{\alpha-1} q_3)-(4 D^{\alpha-1} q_3 - 4 P_{3\alpha} -1)
\\ \label{q3dot}&=& 4P_{2\alpha} + 6 P_{3\alpha} - 4 D^{\alpha-1} q_1 - 8  D^{\alpha-1} q_3 + 1.
\eea
\bea
\nn
\Dot{P}_{1\alpha}&=& \{P_{1\alpha},H_T\}_\alpha = \{P_{1\alpha},D^{\alpha-1} q_1\}_\alpha + (4 D^{\alpha-1} q_3 - 4 P_{3\alpha} -1)\{P_{1\alpha},D^{\alpha-1} q_1\}_\alpha
\\\nn&=&-1-(4 D^{\alpha-1} q_3 - 4 P_{3\alpha} -1)
\\&=&4 P_{3\alpha}-4 D^{\alpha-1} q_3.
\eea
\bea
\Dot{P}_{2\alpha}&=& \{P_{2\alpha},H_T\}_\alpha = \{P_{1\alpha},D^{\alpha-1} q_1\}_\alpha =-1.
\eea
\bea
\nn
\Dot{P}_{3\alpha}&=& \{P_{3\alpha},H_T\}_\alpha = \{P_{3\alpha},(D^{\alpha-1} q_3)^2\}_\alpha 
\\\nn&-&4 \{P_{3\alpha},D^{\alpha-1} q_3\}_\alpha (P_{2\alpha}+P_{3\alpha} -D^{\alpha-1} q_1 - D^{\alpha-1} q_3)
\\\nn&+&(4 D^{\alpha-1} q_3-P_{3\alpha} -1 ) \{P_{3\alpha},D^{\alpha-1} q_3\}_\alpha 
\\\nn&=&-2 D^{\alpha-1} q_3 +4(P_{2\alpha}+P_{3\alpha}-D^{\alpha-1} q_1- D^{\alpha-1} q_3)-4 D^{\alpha-1} q_3+ 4 P_{3\alpha}+1
\\ \label{p3dot}&=&-10 D^{\alpha-1} q_3 + 4 P_{2\alpha}+8 P_{3\alpha}-4 D^{\alpha-1} q_1+1.
\eea
Taking the total time derivative of eq.\eqref{q1dot} we arrive 
\be
D^{\alpha+1} q_1 = \Dot{P}_{1\alpha}   = 4 P_{3\alpha}- 4D^{\alpha-1} q_3.
\ee
Making use of the primary constraint, we get 
\be
\label{q1eq}
D^{\alpha+1} q_1 =2 P_{1\alpha}+2.
\ee
Again using the equation of motion \eqref{q1dot}, we have 
\be
D^{\alpha+1} q_1 -2 D^{\alpha} q_1=2.
\ee
One may using the conformable operator we can rewrite the  above equation in the following form 
\be
\Ddot{q_1} - \Dot{q_1}=2 t^{\alpha-1}.
\ee
This equation can be written as 
\bea
\Dot{y}-2y=2t^{\alpha-1},
\eea
where $y=\Dot{q}_1$. This is a non-homogeneous first-order ordinary differential equation. One may solve it to get 
\bea
y(t)=2e^{2t} \int e^{-2t}t^{\alpha-1}dt +Ae^{2t}.
\eea
Using the incomplete gamma function
\bea
\label{int1}
 \Gamma(s,x)=\int_x^\infty  t^{\alpha-1} e^{-t}  dt.
\eea
We get 
\bea
y(t)=-2^{1-\alpha}  e^{2t}  \Gamma(\alpha,2t) + A e^{2t}.
\eea
Thus,
\bea
\label{q1}
q_1 =\int y(t) dt=- 2^{1-\alpha} \int  e^{2t}  \Gamma(\alpha,2t) dt + \frac{A}{2} e^{2t}+B,
\eea
after simple calculations, we get 
\bea
\label{q1sol}
q_1 =- 2^{-\alpha} e^{2t}\Gamma(\alpha,2t)- \frac{t^\alpha}{\alpha} + \frac{A}{2} e^{2t}+B.
\eea
\begin{figure}[htb!]
    \centering
    \includegraphics[scale=0.9]{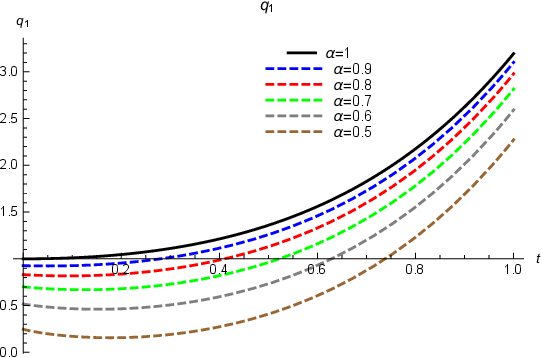}
    \caption{The coordinate $q_1$  at different values of $\alpha$. }
    \label{fig:my_label}
\end{figure}
For $\alpha=1$, we find 
\bea
\nn
q_1 &=&- 2^{-1} e^{2t}\Gamma(1,2t)- t + \frac{A}{2} e^{2t}+B
\\&=&B-\frac{1}{2}- t +\frac{A}{2} e^{2t}
\eea
This is in agreement with ref \cite{rabei1992hamilton}.\\
Taking the total time derivative of eq.\eqref{q2} we have 
\be
\label{eq36}
D^{\alpha+1} q_2 = 3 \Dot{P}_{3\alpha}-4 D^\alpha q_3.
\ee
Inserting eq.\eqref{q3dot} and eq.\eqref{p3dot} we obtain 
\be
\label{46}
D^{\alpha+1} q_2  = 4(2 P_{3\alpha}-2 D^{\alpha-1} q_3).
\ee
Using the conformable secondary constraint we can rewrite eq.\eqref{46} as
\be
\label{q2d}
D^{\alpha+1} q_2  = 4D^{\alpha} q_1 + 4.
\ee
Using the definition of conformable derivative eq.\eqref{q2d} can be written as 
\bea
\Ddot{q}_2=4\Dot{q}_1 + 4t^{\alpha-1}.
\eea
By integrating, we have 
\bea
\Dot{q}_2=4q_1 + 4\frac{t^{\alpha}}{\alpha}+C_1.
\eea
Inserting the solution of $q_1$, we get 
\bea
\nn
 \Dot{q}_2 = - 2^{2-\alpha} e^{2t}\Gamma(\alpha,2t) + 2A e^{2t}+C_2.
\eea
Thus, one may obtain $q_2$ as follows
\bea
\label{q2sol}
q_2= -2^{1-\alpha}e^{2t}\Gamma(\alpha,2t)-2 \frac{t^\alpha}{\alpha}  + A e^{2t}+C_2 t+C_3.
\eea
\newpage
\begin{figure}[htb!]
    \centering
    \includegraphics[scale=0.9]{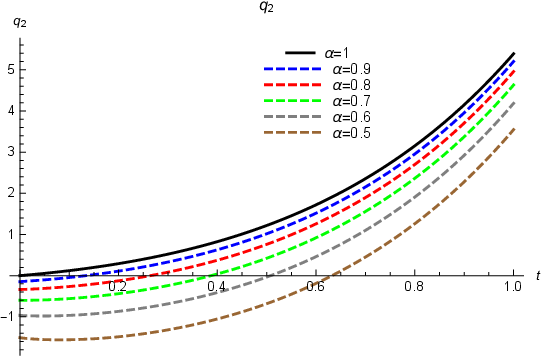}
    \caption{The coordinate $q_2$  at different values of $\alpha$. }
    \label{fig:my_label}
\end{figure}
Again, when  $\alpha=1$, we arrive to the same result in ref \cite{rabei1992hamilton}.\\
Now, making use of the primary constraint \eqref{prim cons}, one may write the equation of motion \eqref{q3dot} as 
\bea
D^{\alpha+1} q_3 +2 D^{\alpha} q_3 = \frac{1}{2} D^{\alpha+1} q_2  . 
\eea
Making use of eq.\eqref{q2sol} and the definition of conformable derivative we get 
\bea
\Ddot{q}_3+2 \Dot{q}_3 = -2^{2-\alpha}e^{2t}\Gamma(\alpha,2t)+ 2 t^{\alpha-1}+2A e^{2t},
\eea
one may solve this equation to find 
\bea
q_3=\frac{e^{-2t}\Gamma(\alpha,-2t)}{2^{\alpha+1}(-1)^\alpha}-\frac{e^{2t}}{2^{1+\alpha}}\Gamma(\alpha,2t)+\frac{A}{4}   e^{2t} - \frac{B}{2}   e^{-2t}+C_4.
\eea
\begin{figure}[htb!]
    \centering
    \includegraphics[scale=0.84]{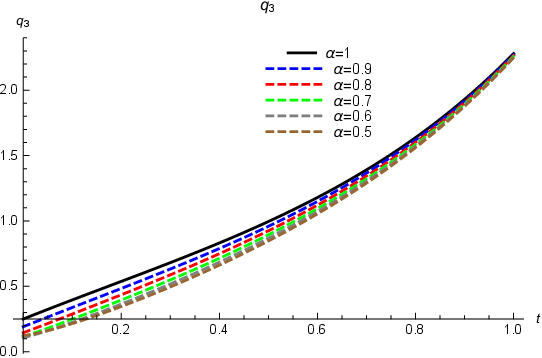}
    \caption{The coordinate $q_3$  at different values of $\alpha$. }
    \label{fig:my_label}
\end{figure}\newpage
When $\alpha=1$, this solution goes to the result obtained in ref \cite{rabei1992hamilton}.
\section{Conclusions}
The Constrained systems which contain conformable derivative  is discussed using the Dirac theory, we see that the Dirac theory is developed to be usable to singular systems containing Conformable order. The equations of motion are written using the conformable Poisson bracket and solved for an illustrative example to obtain the coordinates as functions of time. It is observed that the general solutions of the conformable singular system of second-class constraints go to the results obtained in ref \cite{rabei1992hamilton}  for the traditional second-class constraint when $\alpha=1$. In addition, we draw the general solutions $q_1(t), q_2(t)$, and $q_3(t)$ for different values of $\alpha$, and we have shown that the curve gradually goes to $\alpha=1$. In other words, the curve gradually goes to that of traditional second-class constraints \cite{rabei1992hamilton}. 
\subsection*{Acknowledgement} The authors are grateful to Al al-Bayt University (AABU) and Çankaya University. The first author gratefully acknowledges the AABU for financial and technical support.
\bibliography{ref} 
\bibliographystyle{apalike}
\end{document}